\documentclass[twocolumn,twoside,slac_two]{revtex4}
\usepackage{graphicx}
\usepackage{fancyhdr}
\begin{document}

\title{Study of Mrk~501 above 60~GeV with CELESTE}

%

\author{E. Brion, for the CELESTE collaboration}
\affiliation{CENBG, Chemin du Solarium, Le Haut-Vigneau, BP 120, 33175 Gradignan Cedex, France}

\begin{abstract}
The CELESTE atmospheric Cherenkov detector, running until June 2004 at the Th\'emis solar facility, has taken data on compact sources such as pulsars and blazars. We will take stock of the experiment, in particular regarding the latest improvements of the detector simulation and data analysis. These changes provide us with a new analysis of old data with smaller uncertainties. We present here the evidence for a weak signal from Mrk~501 in 2000-2001.
\end{abstract}

\maketitle

\thispagestyle{fancy}


\section{Introduction}

CELESTE (Cherenkov low energy sampling and timing experiment) was a Cherenkov experiment u\-sing the heliostats of the former \'Electricit\'e de France solar plant in the French Pyrenees at the Th\'emis site. It detected Cherenkov light from electromagnetic sho\-wers produced in the atmosphere by the $\gamma$-rays coming from high energy astrophysical sources. The light is reflected to secondary optics and photomultipliers installed at the top of the tower. Finally it is sampled to be analysed~\cite{Pare}.

Two states of the experiment have to be distinguished to classify CELESTE data. During the first one (between September 1999 and June 2001), 40~heliostats were used with two types of pointing (single pointing: all heliostats at 11~km, double pointing: half at 11~km, half at 25~km), and during the se\-cond one (between September 2001 and June 2004), 53~heliostats poin\-ting at 11~km were used (of which 12 veto heliostats aiming wide for proton rejection). An ana\-lysis improvement has been made on the last data with 53~heliostats taking the Crab (Nebula) data as re\-fe\-ren\-ce, since the source is bright and stable. The new analysis variable provided a sensitivity of 5.1~$\sigma/\sqrt{\mathrm{h}}$ on 5.3~h data, whereas the old data analysis gave 2.0 and 3.4~$\sigma/\sqrt{\mathrm{h}}$ for single and double pointing~\cite{MdN}. The new analysis has been tested on old Crab data with 40~heliostats and gives better sensitivity as will be shown.

CELESTE has taken 40~heliostats data on the blazar Mrk~501. The old analysis gave 2.5~$\sigma$~\cite{LeGallou}. We present here the results obtained with the new ana\-lysis. Mrk~501 is a source interest: it is one of the brightest and closest ($z = 0.034$) extragalactic X-ray and TeV sources. A detection with CELESTE could help to understand the emission processes.

First, the new analysis will be explained with the results obtained on the Crab data. Then, the data selection and analysis, and finally the lightcurve of Mrk~501 will be presented.

\section{Analysis improvement}

\begin{figure}[h!]
  \centering\includegraphics[width=0.90\columnwidth]{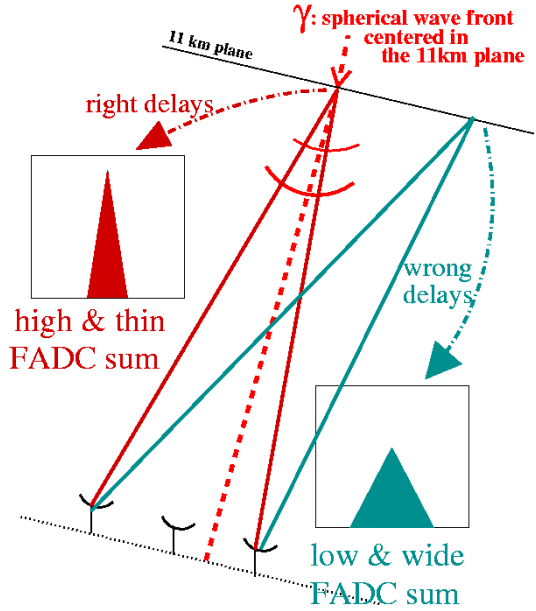}
  \caption{The shape of the FADC sum depends on the position of the center of the wavefront we assume~\cite{Manseri}.}
   \label{fig:FADC}
\end{figure}

\begin{figure}[h!]
  \centering\includegraphics[width=\columnwidth]{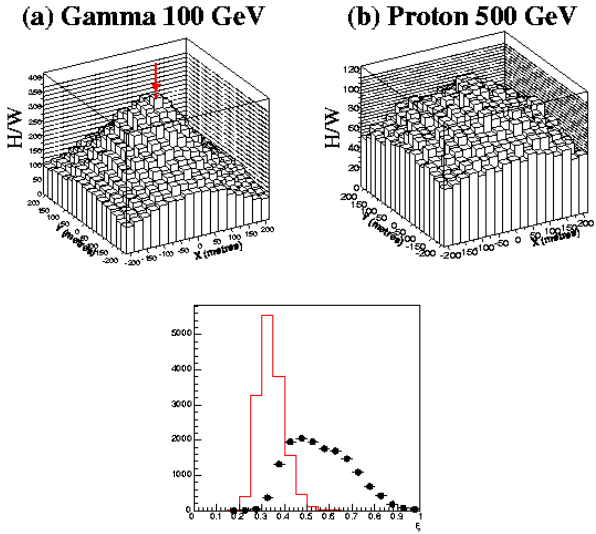}
  \caption{Height-over-width ratio (H/W) computed for discrete positions in the 11~km plane (a) for a 100 GeV $\gamma$-ray and (b) for a 500 GeV proton. (c) Distribution of the $\xi$ parameter for OFF data (filled black circles) and $\gamma$-rays simulation (red line).}
   \label{fig:Timing}
\end{figure}

A new method was found to exploit the FADC information to reject the hadronic background~\cite{Bruel}. To detect low energy $\gamma$-rays, we use the sum of the individual digitized signals to increase the signal-to-background ratio (figure~\ref{fig:FADC}). The summation includes a correction for the sphericity of the wavefront, assumed to be centered in the 11~km plane. Assuming a wrong position for this center (impact parameter) broadens the sum: the height-over-width ratio $(H/W)$ decreases. We compute $(H/W)$ repeatedly for a grid of different assumed positions. The impact parameter is taken to be the position for which the $(H/W)$ ratio is maximum, denoted by $(H/W)_\mathrm{max}$. This is valid for $\gamma$-rays (figure~\ref{fig:Timing}~(a)) but not for protons arriving isotropically and for which the wavefront is not sphe\-rical (fi\-gu\-re~\ref{fig:Timing}~(b)). A measurement of the flatness of these 2D-distributions is given by the following estimator:
$$\xi = \mathrm{average}\,\left(\frac{\displaystyle(H/W)_{200\,\mathrm{m}}}{\displaystyle(H/W)_\mathrm{max}}\right)_{\mbox{\small over 24 positions}},$$
where $(H/W)_{200\,\mathrm{m}}$ is the average of $(H/W)$ over 24 positions along a ring 200~m from the maximum position. For $\gamma$-rays there is a clear maximum and this estimator takes low values. For hadrons, it is usually larger (figure~\ref{fig:Timing}~(c)).

\begin{figure}[h!]
  \centering\includegraphics[width=0.60\columnwidth]{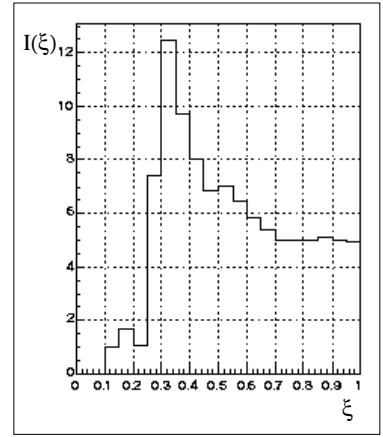}
  \caption{Significance $\sigma$ on the 53~heliostat Crab data integrating $\xi$ by the left: $\mathrm{I(\xi)} = \int_0^\xi\sigma\,\mathrm{d}\xi$. The maximum of the signal is obtained for $\xi<0.35$.}
   \label{fig:TotQFLXsi}
\end{figure}

To improve the signal quality the cut was optimized on the Crab data. For the data with 53~heliostats, the applied cut is $\xi<0.35$ (figure~\ref{fig:TotQFLXsi}) which is the biggest contribution to an improvement in Crab sensitivity to 5.1~$\sigma/\sqrt{\mathrm{h}}$. The detection is stable. 40~heliostat Crab data were also analysed with this variable. Since this estimator is very sensitive, the chosen cut is diffe\-rent: optimizing the cut on 40~heliostats with single poin\-ting gives $\xi<0.30$ ($\xi<0.35$ for double poin\-ting) which is not yet understood. Simulations are in preparation to confirm (or not) this cut. The results are presented in table~\ref{tab:Crab}.

\begin{table}[h!]
\begin{center}
\resizebox{\hsize}{!}{
\begin{tabular}{|l|c|c|c|c|c|}
\hline
Data set & Applied & Significance & Duration & Sensitivity & Flux\\
& cuts & $\sigma$ & [h] & [$\sigma/\sqrt{\mathrm{h}}$] & [$\gamma$/min]\\
\hline \hline
40 heliostats & $\xi<0.30$ & 13.0 & 8.4 & 4.5 & $2.9\pm0.2$\\
single pointing & & & & &\\
\hline
40 heliostats & $\xi<0.35$ & 11.9 & 8.4 & 4.1 & $5.3\pm0.5$\\
single pointing & & & & &\\
\hline
40 heliostats & $\xi<0.35$ & 9.7 & 5.7 & 4.1 & $5.0\pm0.5$\\
double pointing & & & & &\\
\hline
53 heliostats & $\xi<0.35$ & 12.5 & 5.3 & 5.4 & $4.5\pm0.4$\\
with veto& & & & &\\
\hline
53 heliostats & $\xi<0.35$ & 11.6 & 5.3 & 5.1 & $3.5\pm0.3$\\
with veto& no veto & & & &\\
\hline
\end{tabular}
}
\caption{Sensitivity and flux corrected from hour angle efficiency for the different data sets on Crab. All data have an absolute hour angle less than 1.5~h.}
\label{tab:Crab}
\end{center}
\end{table}

\section{Data}

CELESTE took data on Mrk~501 during 2000 and 2001 with 40~heliostats. The previous analysis giving 2.5~$\sigma$, we hope to improve this result with the new analysis.

\subsection{Data selection}

\begin{figure}[h!]
  \centering\includegraphics[width=\columnwidth]{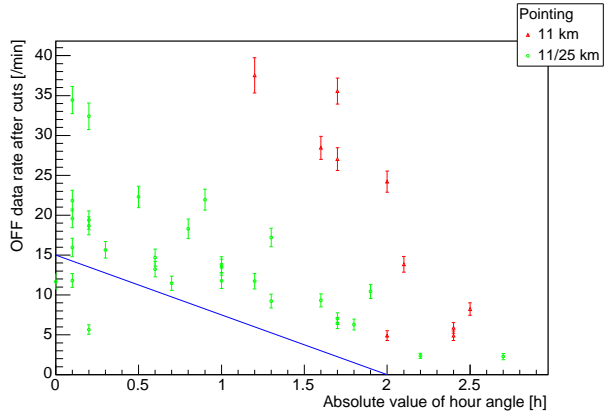}
  \caption{Mrk~501 OFF data rate after analysis cuts as a function of hour angle (typical raw trigger rate is 20~Hz). Data are taken with 40~heliostats single pointing (green circles) and double pointing (red triangles).}
   \label{fig:TxCut}
\end{figure}

Data selection is stricter than in~\cite{LeGallou}. We keep data with stable currents, small ON$-$OFF current and trigger rate differences and number of events $\geq10,000$. Finally, after analysis cuts, the OFF data rate should only depend on the hour angle. Low rates at small hour angle are an indication for bad atmospheric conditions. Thus, we require that they be $\geq-7.5\times|\mathrm{ah}|+15$ (figure~\ref{fig:TxCut}). There was no cut applied on hour angle.

\subsection{Data set and analysis}

After selection, 37~pairs (10.9~h) remain: 10~pairs were taken with single pointing at 11~km and 27 with double pointing at 11/25~km. All pairs are kept together to have more statistics. In terms of years, we have 31~pairs during 2000 and 6~pairs during 2001.

\begin{figure}[h!]
  \centering\includegraphics[width=0.80\columnwidth]{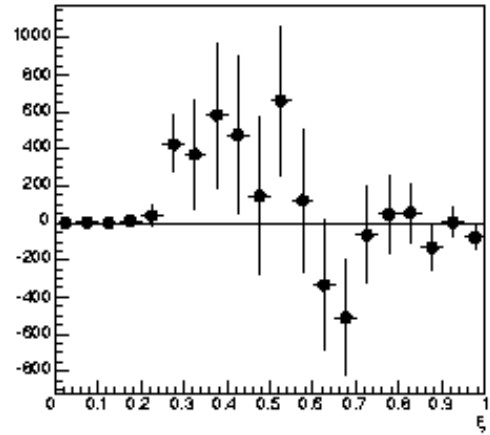}
  \caption{ON$-$OFF $\xi$ distribution for the 40~heliostat Mrk~501 data (single and double pointing).}
   \label{fig:XsiMrk501}
\end{figure}

The results of the new analysis are presented in table~\ref{tab:Mrk501}. Since 40~heliostat Crab data with single and double pointing don't give the same cut to apply on $\xi$, the difficulty here is to choose the cut (figure~\ref{fig:XsiMrk501}). If we apply $\xi<0.30$, we get 2.9~$\sigma$ on all the data (3.1~$\sigma$ in 2000, 0.0~$\sigma$ in 2001). If we apply $\xi<0.35$, we get again 2.5~$\sigma$ on all the data like with the old analysis (2.4~$\sigma$ in 2000, 0.7~$\sigma$ in 2001).

\begin{figure}[h!]
  \centering\includegraphics[width=0.80\columnwidth]{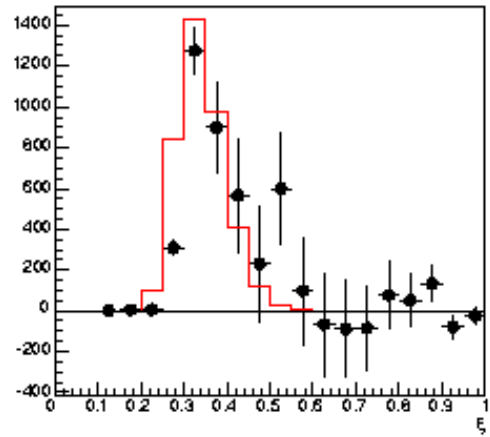}
  \caption{ON$-$OFF $\xi$ distribution for the 53~heliostat Crab data (filled black circles) compared to simulations on Crab normalized to the data (red line).}
   \label{fig:XsiCrab}
\end{figure}

\begin{figure}[h!]
  \centering\includegraphics[width=0.80\columnwidth]{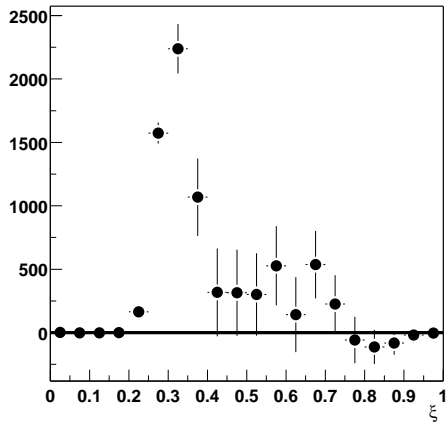}
  \caption{ON$-$OFF $\xi$ distribution for the 53~heliostat Mrk~421 data~\cite{Bruel}.}
   \label{fig:XsiMrk421}
\end{figure}

The chosen cut to present the lightcurve hereafter is $\xi<0.30$, because we have the assumption that the distribution of the $\xi$ variable depends on the declination (owing to its construction). Figure~\ref{fig:XsiCrab} shows this distribution for the 53~heliostat Crab data. In comparison, we can see in figure~\ref{fig:XsiMrk421} the same distribution for 53~heliostat Mrk~421 data, which has the same declination as Mrk~501. The distribution is shifted towards the left.

\begin{table}[h!]
\begin{center}
\resizebox{\hsize}{!}{
\begin{tabular}{|l|c|c|c|c|c|c|}
\hline
Data set & Applied cuts & ON & OFF & ON$-$OFF & Significance & Signal-to-noise\\
& & & & & $\sigma$ & ratio [\%]\\
\hline \hline
All data & raw trigger & 789,648 & 787,452 & 2,195 & 1.5 & 0.3\\
& $\xi<0.30$ & 10,556 & 10,070 & 486 & 2.9 & 4.8\\
& $\xi<0.35$ & 44,218 & 43,363 & 856 & 2.5 & 2.0\\
\hline
Year 2000 & raw trigger & 666,704 & 663,474 & 3,230 & 2.3 & 0.5\\
& $\xi<0.30$ & 9,424 & 8,940 & 485 & 3.1 & 5.4\\
& $\xi<0.35$ & 38,753 & 37,977 & 776 & 2.4 & 2.0\\
\hline
Year 2001 & raw trigger & 122,944 & 123,978 & $-1,035$ & $-1.9$ & --\\
& $\xi<0.30$ & 1,132 & 1,130 & 1.8 & 0.0 & 0.2\\
& $\xi<0.35$ & 5,465 & 5,385 & 80 & 0.7 & 1.5\\
\hline
\end{tabular}
}
\caption{Mrk~501 number of events (40~heliostats single and double pointing) after different cuts. Note that the number of events is rounded.}
\label{tab:Mrk501}
\end{center}
\end{table}

A work is in preparation that compares thoses distributions for Crab, Mrk~421 and Mrk~501 40~heliostat data with simulations. Since we have a lot of data on Mrk~421, we will be able to optimize our cut on this source and apply it to Mrk~501.

\subsection{Seasonal changes}

CELESTE raw trigger rates are typically 25~Hz du\-ring the Winter. In early Fall and in Spring, rates fall by a factor of two. The LIDAR on site showed that increased atmospheric extinction due to aerosols is a major cause~\cite{Bussons}. Unfortunately, it started to get data in December 2001 after these data on Mrk~501. We looked at the possible influence of this effect on the ON$-$OFF excess, but nothing was highlighted: for trigger rate $\leq15$~Hz, 20~pairs give 2.7~$\sigma$ during 6.0~h, and for trigger rate $>15$~Hz, 17~pairs give 1.5~$\sigma$ during 4.9~h.

\section{Lightcurve}

\begin{figure}[h!]
  \centering\includegraphics[width=\columnwidth]{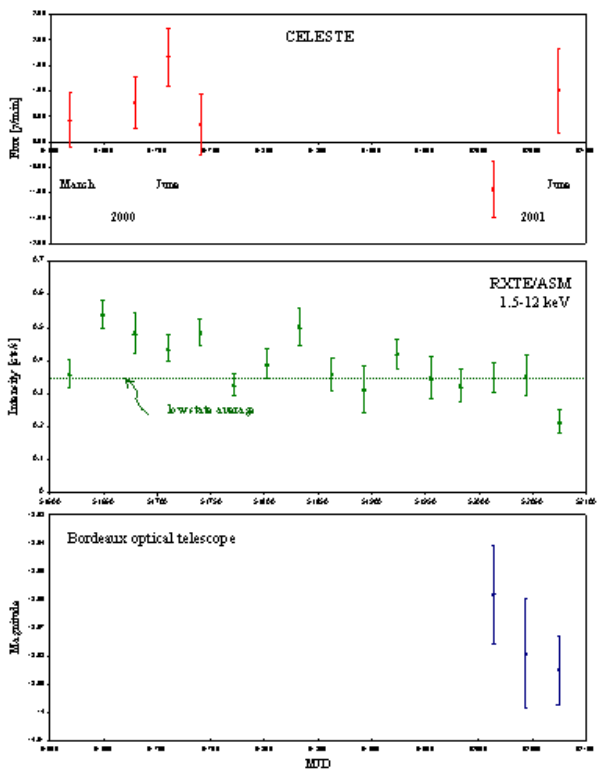}
  \caption{Mrk~501 lightcurves seen by CELESTE above 60~GeV in 2000-2001, RXTE/ASM in X-rays~\cite{ASM}, and the optical telescope of Bordeaux~\cite{Charlot}. The CELESTE lightcurve is uncorrected for efficiency in hour angle, nor for different atmospheric extinctions.}
   \label{fig:LightCurve}
\end{figure}

We present in figure~\ref{fig:LightCurve} the lightcurve obtained by CELESTE for monthly averages with the chosen cut $\xi<0.30$. Note that Mrk~501 was active in X-rays du\-ring 2000 and quiet during 2001~\cite{Xue}. During the year 2000, the cut to be confirmed, we get an ON$-$OFF excess of 3.1~$\sigma$ during 9.1~h.

\section{Conclusion}

Before the LIDAR study, the acceptance was unknown for data with trigger rates less than 15 Hz. We now have more confidence in the 3.1~$\sigma$ obtained for the year 2000 and in our acceptance. Therefore, if this result is confirmed, we will interpret this excess as a weak signal and calculate a flux. The acceptance will include the degraded atmospheric extinction.

If we interpret this excess as a signal, and under the (false) assumption that we have the same detector acceptance and spectrum for Mrk~501 as for the Crab, we get one fourth of the Crab flux. This is comparable with the results presented in~\cite{Tavecchio} (data on April 29, 1998 in X-rays similar to Spring 2000).


\bigskip 

\end{document}